\documentclass[12pt,a4paper]{article}
\usepackage{amssymb}
\newcommand{\R}{\mathbb{R}}
\newcommand{\N}{\mathbb{N}}
\newcommand{\epn}{\epsilon}
\newcommand{\p}{\partial}
\newcommand{\nn}{\nonumber\\}
\newcommand{\hatg}{\hat{\gamma}}

\newcommand{\hi}{\hat{i}}
\newcommand{\hj}{\hat{j}}
\newcommand{\hk}{\hat{k}}
\newcommand{\hl}{\hat{l}}

\newcommand{\hM}{\hat{M}}
\newcommand{\hN}{\hat{N}}
\newcommand{\hP}{\hat{P}}
\newcommand{\hA}{\hat{A}}
\newcommand{\hC}{\hat{C}}
\newcommand{\elpl}{l_{11}}
\newcommand{\hV}{\hat{V}}
\newcommand{\hh}{\hat{h}}

\newcommand{\GO}{\mathring{G}}
\newcommand{\hPi}{\hat{\Pi}}
\newcommand{\hE}{\hat{E}}
\newcommand{\hG}{G}
\newcommand{\ga}{g}
\newcommand{\calD}{\mathcal{D}}

\newcommand{\fdg}{\bar{\gamma}}
\newcommand{\hmu}{\mu}
\newcommand{\hnu}{\nu}
\newcommand{\hrho}{\rho}
\newcommand{\hsig}{\sigma}

\newcommand{\hze}{\hat{\zeta}}
\newcommand{\hld}{\hat{\lambda}}
\newcommand{\pc}{{\phi_0}}

\newcommand{\gstg}{\gamma}
\newcommand{\BI}[1]{{\mathring{#1}}}

\newcommand{\ha}{\hat{a}}
\newcommand{\hb}{\hat{b}}
\newcommand{\estg}{\tilde{e}}
\newcommand{\cg}{\hat{c}}
\newcommand{\bg}{\hat{b}}
\newcommand{\toDDR}{\mathop{\to}\limits_{DDR}}

\title{Dilaton coupling revisited}
\author{
{\sc Shozo Uehara}\footnote{e-mail:
	uehara@is.utsunomiya-u.ac.jp}\vspace{4mm}\\
{\sl Department of Information Systems Science, Utsunomiya
	University,}\\
{\sl Utsunomiya 321-8585, Japan}}
\date{}
\makeatletter

 \@addtoreset{equation}{section}
\renewcommand{\thefigure}{\@arabic\c@figure}
\makeatother
\begin{document}
\maketitle

\begin{abstract}
We reinvestigate the dilaton coupling in the string theory, which
comes from a wrapped membrane.
The ghost number anomaly associated with the string worldsheet
diffeomorphism is shown to induce the dilaton coupling.
\end{abstract}

\newpage
\section{Introduction}
Supermembrane  in eleven dimensions \cite{BST} is an important object
in M-theory and the relation with superstring is well known.
In fact, through the double dimensional reduction \cite{DHIS} the
wrapped supermembrane on $\R^{10}\times S^1$ is reduced to type IIA
superstring on $\R^{10}$.
Furthermore, it was explicitly shown that the ($p,q$)-string
\cite{Sch,W} in type IIB theory is obtained from the wrapped
supermembrane on the $T^2$-compactified target space through the
double dimensional reduction \cite{OUY,OUY2,OUY3}.
This also indicates that the duality in type IIB theory is naturally
understood in M-theory \cite{Sch,Asp}.

Recently, the coupling of the string worldsheet Euler character $\chi$
to the dilaton $\phi$ has been studied in M-theory \cite{BP}.
It was presented that the $\chi\phi$-term in type IIA theory
arises from the measure of the membrane partition function.
This may also indicate that the membrane explains the properties of
string theory.
In this paper we reinvestigate the membrane origin of the dilaton
term by the Fujikawa method \cite{F,F1} and we will see that the
path integral measure leads to the coupling of the dilaton $\phi$ to
the Ricci scalar $R^{(2)}$ of the string worldsheet metric $\gstg$
\begin{equation}
 \int_{\Sigma} d^2\sigma \sqrt{-\gstg}\,R^{(2)}\phi\,.
\end{equation}

This paper is organized as follows.
In the next section, we deduce the string action from the wrapped
supermembrane by the double dimensional reduction.
In section \ref{S:D}, we seek for the dilation coupling with the
partition function of the reduced string action.
The final section is devoted to discussion.

\section{Double dimensional reduction}\label{S:F}
The action of a supermembrane coupled to an eleven-dimensional
supergravity background is given by \cite{BST}
\begin{equation}
 S=T\int\!d^3\sigma\Biggl[-\frac{1}{2}\,\sqrt{-\hatg}\,\hatg^{\hi\hj}
	\,\hPi_{\hi}^{~A}\hPi_{\hj}^{~B}\,\eta_{AB}
    +\frac{1}{2}\,\sqrt{-\hatg}
    -\frac{1}{3!}\,\epn^{\hi\hj\hk}\,\p_{\hi}Z^{\hM}\p_{\hj}Z^{\hN}
	\p_{\hk} Z^{\hP}\hC_{\hP\hN\hM}\Biggr]\label{eq:SMac},
\end{equation}
where
\begin{equation}
  \hPi_{\hi}^{~\hA}=\p_{\hi} Z^{\hM}\hE_{\hM}^{~\hA}\,,
\end{equation}
$T$ is the tension of the supermembrane,\footnote{The
eleven-dimensional Planck length $l_{11}$ is defined by
$T=(2\pi)^{-2}l_{11}^{-3}$.} $\hC_{\hM\hN\hP}(Z)$ is the super
three-form, $\hatg_{\hi\hj}\ (\hi,\hj=0,1,2)$ is the
worldvolume metric, $\hatg=\det\hatg_{\hi\hj}$, and the target space
is a supermanifold with the superspace coordinates
$Z^{\hM}=(X^M,\theta^{\alpha})~(M=0,\cdots,10,~\alpha=1,\cdots,32)$.
Furthermore, with the tangent superspace index $\hA=(A,a)$,
$\hE_{\hM}^{~\hA}$ is the supervielbein and $\eta_{AB}$ is the
tangent space metric in eleven dimensions.
The mass dimension of the worldvolume parameter $\sigma^{\hi}$ and
eleven-dimensional background fields ($G_{MN}$, $\hC_{MNP}$) is
$0$, while that of the worldvolume metric $\hatg_{\hi\hj}$ is $-2$.

We shall investigate the origin of the $R^{(2)}\phi$-term in the
dimensionally reduced string theory.
In string theory, the anomaly of the reparametrization ghost number
current gives $\sqrt{-\gstg}\,R^{(2)}$, which is (a local version of)
the Riemann-Roch theorem, and hence, similarly to Ref.\cite{BP}, we
may focus on the bosonic degrees of freedom to investigate
the $R^{(2)}\phi$-term from the supermembrane theory.

The bosonic background fields are included in the superfields as
\begin{equation}
  \hE_M^{~A}(Z)\bigg|_{\mathrm{fermions}=0}
	= \hat{e}_M^{~A}(X)\,,\quad
  \hC_{MNP}(Z)\bigg|_{\mathrm{fermions}=0}= \hA_{MNP}(X)\,.
\end{equation}
Then, the action (\ref{eq:SMac}) is reduced to
\begin{eqnarray}
  S&=&T\int\!d^3\sigma\Biggl[-\frac{1}{2}\sqrt{-\hatg}\,
	\hatg^{\hi\hj}\, \p_{\hi} X^M\p_{\hj} X^N\hG_{MN}(X)
	+\frac{1}{2}\sqrt{-\hatg}\nn
  &&\hspace{10ex}{}-\frac{1}{3!}\,\epn^{\hi\hj\hk}\,\p_{\hi}X^M
    \p_{\hj}X^N\p_{\hk}X^P \hA_{PNM}(X)\Biggr]\label{eq:Mac}.
\end{eqnarray}
Note that the variation w.r.t.\ $\hatg_{\hi\hj}$ yields the induced
metric,
\begin{equation}
  \hat\gamma_{\hi\hj}=\p_{\hi}X^M \p_{\hj}X^N\hG_{MN}(X)
	\equiv\hG_{\hi\hj}\,,\label{eq:IM}
\end{equation}
and plugging it back into the original action leads to the one in the
Nambu-Goto form
\begin{equation}
 S^{\mathrm{(NG)}}
  =T\int\!d^3\sigma\,\Biggl[-\sqrt{-\det\hG_{\hi\hj}}
    -\frac{1}{3!}\,\epn^{\hi\hj\hk}\p_{\hi}X^M
    \p_{\hj}X^N \p_{\hk}X^P\hA_{PNM}(X)\Biggr]\,.\label{eq:Mac1}
\end{equation}

In Ref.\cite{BP}, they examined the membrane partition function
\begin{equation}
 Z=\sum_{\mathrm{topologies}}
  \int\frac{\calD X\calD\hatg}{\mathrm{Vol(Diff}_0)}\,e^{-S}\,,
  \label{eq:Zbp}
\end{equation}
under the assumption that the membrane was wrapped once around the
$S^1$-compactified 11th direction of the target space and they
truncated to the zero-mode sector of the circle.
That is, the worldvolume topology was assumed to be $S^1\times\Sigma$
with $\Sigma$ being some Riemann surface and the target space topology
was $M^{10}\times S^1$ with $S^1$ being the 11th direction of M-theory
or the M-theory circle, and the double dimensional reduction
\cite{DHIS} was applied. They fixed some variables
\begin{equation}
  \sigma^2=X^{11}\,,\quad \hatg_{22}=1\,,
  \quad\hatg_{02}=\hatg_{12}=0\,,\quad G_{1010}=1\,,
\end{equation}
and analyzed the path integral measure of the partition function
(\ref{eq:Zbp}).
They found that the norm of the variation of the worldvolume metric
leads to the relation
\begin{equation}
  ||\delta\hatg_{ij}||=\sqrt{R_{11}}\, ||\delta\gstg_{ij}||,\label{eq:GGr}
\end{equation}
where $R_{11}$ is the radius of the M-theory circle.
Then, (\ref{eq:GGr}) is followed by the relation between the moduli
space (or the conformal Killing vectors) measures of the string and
the dimensionally reduced wrapped membrane, which leads to the
$\chi\phi$-term to the string action.

In this paper, we reinvestigate the partition function (\ref{eq:Zbp}).
Similarly to Ref.\cite{BP}, since we are interested in the coupling
between the dilaton and the worldsheet curvature in the reduced string
theory, we consider the $S^1$-compactified wrapped supermembrane, for
simplicity.
We take the $X^{10}$-directions to be compactified on $S^1$ of radius
$L_1$ and hence the worldvolume of the membrane is at least locally
$\Sigma_{ws}\times S^1$, where  $\Sigma_{ws}$ is a Riemann surface and
$S^1$ is to be parametrized by $\sigma^2$.
Then, we shall represent the wrapping of the supermembrane as
\begin{equation}
 \sqrt{\strut\GO_{1010}}\,X^{10}(\sigma^{\hi}+2\pi\delta_{\hi2})
  =2\pi w_1L_1 +\sqrt{\strut\GO_{1010}}\,X^{10}(\sigma^{\hi})\,,
  \label{eq:BC}
\end{equation}
where $\GO_{1010}$ stands for the asymptotic values of the
corresponding component of the target space metric and $w_1\in\N$ is a
wrapping number. For simplicity,  we put $w_1=1$ hereafter.
{}From eq.(\ref{eq:11BG}) we shall see
\begin{equation}
  R_1\equiv \frac{L_1}{\sqrt{\GO_{1010}}}=L_1\,e^{-2\pc/3},
	\label{eq:R1}
\end{equation}
where $\pc$ is the asymptotic constant value
of the type IIA dilaton background and hence the M/IIA-relation or
11d/IIA-SUGRA-relation leads to
\begin{equation}
	R_1=\elpl\,.\label{eq:R1pl}
\end{equation}

We parametrize the worldvolume metric $\hatg_{\hi\hj}$ as
\begin{equation}
  \Bigl(\,\hatg_{\hi\hj}\,\Bigr)
    =\elpl^2 \left(\begin{array}{@{\,}cc@{\,}}
	h^{-1/2}\,\gstg_{ij}+ h V_iV_j & h\,V_i \\
	h\,V_j & h \end{array} \right),\label{eq:WV1A}
\end{equation}
where $\gstg_{ij}$, $h$ and $V_i$ are dimensionless, and we
have\footnote{Note that for an arbitrary worldvolume vector
$\hA_{\hi}$ and a tensor $\hA_{\hi\hj}$,  we have
\begin{eqnarray}
 &&\hA^{\hi}=\hatg^{\hi\hk}\hA_{\hk}\to
	 \hA^{i}=\elpl^{-2}h^{1/2}\gstg^{ik}(\hA_k-V_k\hA_2)\,,\nn
 &&\hA^{\hi\hj}=\hatg^{\hi\hk}\hatg^{\hj\hl}\hA_{\hk\hl}
	\to\hA^{ij}= \elpl^{-4}h\gstg^{ik}\gstg^{jl}(\hA_{kl}
	-V_k\hA_{2l}-\hA_{k2}V_l+V_kV_l\hA_{22})\,.
\end{eqnarray}
Meanwhile, the first two components $\epn^i$ of a parameter
$\epn^{\hi}$ for the worldvolume diffeomorphism can be a
parameter of the worldsheet diffeomorphism because, for example,
($\hmu=0,1,\cdots,9$)
\begin{equation}
  \delta X^M = \epn^{\hi}\p_{\hi}X^M \toDDR
  \delta X^{\hmu} = \epn^{i}\p_i X^{\hmu}.\quad
\end{equation}}
\begin{eqnarray}
 \det (\hatg_{\hi\hj}) &=&\elpl^6  \det (\gstg_{ij})\,,\\
 \Bigl(\,\hatg_{\hi\hj}\,\Bigr)^{-1}&=&\Bigl(\,\hatg^{\hi\hj}\,\Bigr)
    =\elpl^{-2}\left(\begin{array}{@{\,}cc@{\,}}
	h^{1/2}\,\gstg^{ij}& -h^{1/2}\,\gstg^{ik}V_k \\
	-h^{1/2}\,\gstg^{jk}V_k\, &
	h^{-1}+h^{1/2}V_k\gstg^{kl}V_l \end{array}\right).
\end{eqnarray}
Then, the action (\ref{eq:Mac}) is rewritten as
\begin{eqnarray}
 S&=&\frac{T}{2}\int\!d^3\sigma\,\Biggl[
    -\elpl\sqrt{-\gstg}\,h^{1/2}\gstg^{ij}\,(\p_i X\cdot\p_j X
    -\elpl^2\hh\hV_i\hV_j)-\elpl^3\sqrt{-\gstg}\,(h^{-1}\hh-1)\nn
  &&\quad{}-\elpl^3\sqrt{-\gstg}\,h^{1/2}\hh\,
	\gstg^{ij}(V_i-\hV_i)(V_j-\hV_j)
	+\epn^{ij}\,\p_iX^{M}\p_jX^{N}\p_2X^{P}\hA_{MNP}\Biggr],
	\quad\label{eq:Mac0A}
\end{eqnarray}
where
\begin{equation}
 \p_{\hi}X\cdot\p_{\hj}X=\p_{\hi}X^M\p_{\hj}X^N\hG_{MN}\,,\quad
 \hh=\elpl^{-2}\,\p_2X\cdot\p_2X\,,\quad
 \hV_i=\frac{\p_iX\cdot\p_2X}{\p_2X\cdot\p_2X}\,.\label{eq:hhV}
\end{equation}

We shall make a gauge choice for the worldvolume
diffeomorphism.\footnote{Under the diffeomorphism,
the components in (\ref{eq:WV1A}) are transformed as
\begin{eqnarray}
 \delta\gstg_{ij}&=&\epn^{\hi}\p_{\hi}\gstg_{ij}
	+\gstg_{ij}\p_2\epn^2
	+(\gstg_{ij}V_k-\gstg_{jk}V_i-\gstg_{ki}V_j)\p_2\epn^k
	+\gstg_{ik}\p_j\epn^k+\gstg_{jk}\p_i\epn^k,\label{eq:diff3g}\\
 \delta V_i &=&\epn^{\hi}\p_{\hi}V_i+V_j\,\p_i\epn^j
	+(h^{-3/2}\gstg_{ij}-V_iV_j)\,\p_2\epn^j+\p_i\epn^2
	-V_i\p_2\epn^2,\label{eq:diff3v}\\
 \delta h &=&\epn^{\hi}\p_{\hi}h+2h\,\p_2\epn^2 +2hV_i\,\p_2\epn^i.
	\label{eq:diff3h}
\end{eqnarray}}
We adopt the following gauge conditions\footnote{We may adopt
$\p_2X^{10}=R_1$ instead of $X^{10}=R_1\sigma^2$ to see the conformal
transformation explicitly in the dimensionally reduced string
theory;\cite{AKS} however, it does not make an essential difference in
our analysis.}
\begin{equation}
  X^{10}= R_1\,\sigma^2,
	\quad \gstg_{01}=\gstg_{00}+\gstg_{11}=0\,.\label{eq:GF}
\end{equation}
Under the gauge condition (\ref{eq:GF}) the action (\ref{eq:Mac}) or
(\ref{eq:Mac0A}) becomes
\begin{eqnarray}
S_{\mathrm{Gfd}}&=&\frac{T}{2}\int\!d^3\sigma\,\Biggl[
    -\frac{h^{1/2}\eta^{ij}}{\hG_{1010}^{1/2}}\,
	\Bigl\{\elpl\p_i X^{\hmu} \p_jX^{\hnu}g_{\hmu\hnu}
    -2\frac{\hG_{1010}}{\hh}\,\p_i X^{\hmu}\p_j X^{\hnu}
	\p_2X^{\hrho}g_{\hmu\hrho}A_{\hnu}\nn
  &&{}\quad+\frac{\hG_{1010}}{\elpl\hh}\,
	\p_i X^{\hmu}\p_j X^{\hnu}\p_2X^{\hrho}\p_2X^{\hsig}
	\Bigl(A_{\hmu}A_{\hnu}g_{\hrho\hsig}
	-2A_{\hnu}A_{\hsig}g_{\hmu\hrho}
	-\frac{g_{\hmu\hrho}g_{\hnu\hsig}}{\hG_{1010}^{3/2}}\Bigr)
	\Bigr\}\nn
  &&{}+\elpl\epn^{ij}\p_iX^{\hmu}\p_jX^{\hnu}B_{\hmu\hnu}
	+\epn^{ij}\p_iX^{\hmu}\p_jX^{\hnu}
	\p_2X^{\hrho}A_{\hmu\hnu\hrho}\nn
  &&{}-\elpl^3\rho h^{-1}(\hh-h)-\elpl^3h^{1/2}\hh\,
	\eta^{ij}(V_i-\hV_i)(V_j-\hV_j)\Biggr],\label{eq:MACgfd}
\end{eqnarray}
where $\gstg_{ij}$ has been fixed to be the fiducial metric
$\fdg_{ij}$ by the gauge condition (\ref{eq:GF})
\begin{equation}
	\fdg_{ij}\equiv\rho\eta_{ij},\label{eq:FDm}
\end{equation}
and use has been made of the (10+1)-decomposition of the background
fields in  (\ref{eq:11BG}) and (\ref{eq:dhA}).

The FP-determinant $\Delta_{\mathrm{FP}}$, or the Fadeev-Popov ghost
action $S_{\mathrm{FP}}$,  associated with the gauge condition
(\ref{eq:GF}) is given by (see appendix \ref{S:FP02a} and also
\cite{KU})
\begin{eqnarray}
 \Delta_{\mathrm{FP}}(\hatg,X)&=&\int\!\calD\bg_{++}\,
	\calD\bg_{--}\,\calD\cg^+\, \calD\cg^-\,
	\exp\Bigl(iS_{\mathrm{FP}}\Bigr),\\
 S_{\mathrm{FP}}&=&-\frac{i}{2\pi^2}\!\int\!\!d^3\sigma\,h^{1/2}\Bigl\{
	\bg_{--}(\p_+-V_+\p_2)\cg^-
	+\bg_{++}(\p_- -V_-\p_2)\cg^+\Bigr\},\label{eq:MacFP}
\end{eqnarray}
where $\bg_{\pm\pm}$ and $\cg^{\pm}$ are Grassmann odd ghosts with
ghost numbers $-1$ and $+1$, respectively, and
\begin{equation}
 \p_{\pm}=\frac{1}{2}\,(\p_0\pm\p_1),
	\quad V_{\pm}=\frac{1}{2}\,(V_0\pm V_1)\,.
\end{equation}
The total action is given by
\begin{equation}
  S_T=S_{\mathrm{Gfd}}+S_{\mathrm{FP}}.\label{eq:MacT}
\end{equation}

Now that we make the double dimensional reduction by imposing
the conditions to deduce the type IIA string ($\hmu=0,1,\cdots,9$)
\begin{equation}
  \p_2 X^{\hmu}\,(=\p_{\sigma^2} X^{\hmu})=0\,,\quad
	\frac{\p}{\p X^{10}}\,\hG_{MN}
	=\frac{\p}{\p X^{10}}\,\hA_{MNP} =0\,.\label{eq:DDRA}
\end{equation}
By the double dimensional reduction the total  action is reduced to
\begin{eqnarray}
 S_T\toDDR
 S_{st}&=&\frac{T_s}{2}\int\!d^2\sigma
    \Biggl[-\sqrt{\frac{h}{\bar{h}}}\,
    \eta^{ij}\p_i X^{\hmu}\p_j X^{\hnu}g_{\hmu\hnu}
	+\epn^{ij}\p_iX^{\hmu}\p_jX^{\hnu}B_{\hmu\hnu}\nn
  &&{}-\frac{1}{2\pi T_s}\rho\Bigl(\frac{\bar{h}}{h}-1\Bigr)
	-\frac{1}{2\pi T_s}h^{1/2}\bar{h}\,
	\eta^{ij}(V_i-\bar{V}_i)(V_j-\bar{V}_j)\nn
  &&{}-\frac{2i}{\pi T_s}h^{1/2}(b_{--}\p_+c^-
	+b_{++}\p_-c^+)\Biggr],\label{eq:Mac02a}
\end{eqnarray}
where
\begin{eqnarray}
 &&T_s=2\pi R_1T=(2\pi\elpl^2)^{-1},\\
 &&\hh\toDDR\hG_{1010}\equiv\bar{h}\,,\quad
  \hV_i\toDDR\frac{\hG_{\hmu10}\p_iX^{\hmu}}{R_1\hG_{1010}}
	\equiv\bar{V}_i\,,\label{eq:bhV}
\end{eqnarray}
and the variables in $S_{st}$ are understood to be independent of
$\sigma^2$.
Note that $V_i$ can be integrated out straightforwardly, while
the  field equation for $\rho$ is
\begin{equation}
  h=\bar{h}\,.
\end{equation}
As was pointed out in \cite{AKS}, substituting this algebraic
condition back into $S_{st}$ yields a $\rho$-independent action.
However, $\rho$ is not free but is related, for example,  to
$X^{\hmu}$ through the equation of motion for $h$.
The Euclidean action corresponding to $S_{st}$ is
\begin{eqnarray}
 S^{E}_{st}&=&T_s\int\!d^2z\Biggl[\sqrt{\frac{h}{\bar{h}}}\,
    \p_z X^{\hmu}\p_{\bar{z}} X^{\hnu}g_{\hmu\hnu}
	+\p_zX^{\hmu}\p_{\bar{z}}X^{\hnu}B_{\hmu\hnu}
	+\frac{1}{8\pi T_s}\rho\,\Bigl(\frac{\bar{h}}{h}-1\Bigr)\nn
  &&{}+\frac{1}{2\pi T_s}h^{1/2}\bar{h}\,
	(V_z-\bar{V}_z)(V_{\bar{z}}-\bar{V}_{\bar{z}})
    +\frac{1}{2\pi T_s}\,h^{1/2}(b_{zz}\p_{\bar{z}}c^z
      +b_{\bar{z}\bar{z}}\p_zc^{\bar{z}})\Biggr].\label{eq:Mac02aE}
\end{eqnarray}

\section{Dilaton coupling}\label{S:D}
In this section we examine the partition function with the double
dimensional reduced membrane action.
First, we shall make the BRS invariant path integral measure according
to the Fujikawa method \cite{F}.
In this case, the integration variables for the path integral are
obtained as follows:
For a worldvolume scalar $X^{\hmu}$ we have
\begin{eqnarray}
 &&\int\calD\BI{X}^{M}\exp\Biggl[T\elpl^{-2}\int\!d^3\sigma
	\sqrt{-\hatg}\, X^MX^N\hG_{MN}\Biggr]\nn
 &&=\int\!\calD\BI{X}^M\exp\Biggl[T\elpl\!\int\!d^3\sigma\,
    \sqrt{-\gstg}\,\Biggl\{X^{\hmu}X^{\hnu}
	\frac{g_{\hmu\hnu}}{\sqrt{\hG_{1010}}}+\hG_{1010}\Bigl(X^{10}
	+\frac{\hG_{10\hmu}X^{\hmu}}{\hG_{1010}}\Bigr)^2
	\Biggr\}\Biggr].\nn
\end{eqnarray}
This implies (we shall set $\det g_{\hmu\hnu}=-1$)
\begin{equation}
 \calD\BI{X}^{\hmu}=\calD\Bigl((-\gstg/\bar{h})^{1/4}
	X^{\hmu}\Bigr).\label{eq:BIX}
\end{equation}
Meanwhile, we have the following decomposition for a worldvolume
vector,
\begin{equation}
 \elpl^{-2}\hatg_{\hi\hj}A^{\hi}A^{\hj}
	=h^{-1/2}\gstg_{ij}A^iA^j+ h(A^2+A^iV_i)^2\,.
\end{equation}
This implies ($\gstg_{ij}=\estg_i^{~a}\estg_{ja}$, ($\estg_i^{~a}$:
 zweibein))
\begin{equation}
 \calD\BI{A}^{\ha}
  =\calD\Bigl((-\gstg/h)^{1/4}\tilde{e}_i^{~a}A^i\Bigr)\,
	\calD\Bigl((-\gstg h^2)^{1/4}A^2\Bigr)\nn
  =\calD \Bigl((-\gstg)^{1/2}h^{-1/4}A^i\Bigr)\,
	\calD\Bigl((-\gstg)^{1/4} h^{1/2}A^2\Bigr)\,.\label{eq:BinvV0}
\end{equation}
Similarly, for a symmetric tensor we have
\begin{equation}
 \calD\BI{\Phi}_{\ha\hb}=\calD\Bigl((-\gstg)^{-1/4}
	h^{1/2}\Phi_{ij}\Bigr)
	\calD\Bigl(h^{-1/4}\Phi_{2i}\Bigr)\calD\Bigl(
      (-\gstg)^{1/4}h^{-1}\Phi_{22}\Bigr)\,.\label{eq:BinvT0}
\end{equation}

Let us rewrite $S_{st}$ with the path integral variables
$\BI{X}, \BI{b}$ and $\BI{c}$.
We have (cf. (\ref{eq:BIX}), (\ref{eq:BinvV0}), (\ref{eq:BinvT0}))
\begin{eqnarray}
 \BI{X}^{\hmu}&=& \rho^{1/2}\,\bar{h}^{-1/4} X^{\hmu},\nn
 \BI{c}^z&=& \rho\,h^{-1/4}c^z,\nn
 \BI{b}_{zz}&=&\rho^{-1/2}h^{1/2}b_{zz}\,.
\end{eqnarray}
Thus, $S^E_{st}$ becomes
\begin{eqnarray}
 S^E_{st}&=&\int\!d^2z\Biggl[T_s\sqrt{\frac{h}{\bar{h}}}\,\p_z
    \Biggl(\frac{\BI{X}^{\hmu}}{\rho^{1/2}\bar{h}^{-1/4}}
    \Biggr)\p_{\bar{z}}\Biggl(\frac{\BI{X}^{\hnu}}
	{\rho^{1/2}\bar{h}^{-1/4}}\Biggr)g_{\hmu\hnu}\nn
  &&{}+T_s\sqrt{\frac{h}{\bar{h}}}\,\p_z
    \Biggl(\frac{\BI{X}^{\hmu}}{\rho^{1/2}\bar{h}^{-1/4}
	}\Biggr)\p_{\bar{z}}\Biggl(\frac{\BI{X}^{\hnu}}
	{\rho^{1/2}\bar{h}^{-1/4}}\Biggr)B_{\hmu\hnu}
	+\frac{1}{8\pi}\rho\Bigl(\frac{\bar{h}}{h}-1\Bigr)\nn
  &&{}+\frac{1}{2\pi}\,\rho^{1/2}\BI{b}_{zz}\p_{\bar{z}}
	\Bigl(\rho^{-1}h^{1/4}\BI{c}^z\Bigr)
	+\frac{1}{2\pi}\,\rho^{1/2}\BI{b}_{\bar{z}\bar{z}}\p_z
	\Bigl(\rho^{-1}h^{1/4}\BI{c}^{\bar{z}}\Bigr)\Biggr],
\end{eqnarray}
and the partition function is
\begin{equation}
 Z\sim\int\!\calD\BI{X}^{\hmu}\calD\BI{c}^z\calD\BI{c}^{\bar{z}}
	\calD\BI{b}_{zz}\calD\BI{b}_{\bar{z}\bar{z}}\, e^{-S^E_{st}}.
\end{equation}
We find that $\rho$ and also $h$ can be removed from the action
$S^E_{st}$ by putting $h=\bar{h}$ and making the following rescalings,
\begin{eqnarray}
 (\BI{X},\BI{b},\BI{c})&\to&
	(e^{\alpha_c/2}\BI{X},e^{-\alpha_c/2}\BI{b},
	e^{\alpha_c}\BI{c}),\quad
	(\,e^{\alpha_c}=\rho\bar{h}^{-1/2}\,)\label{eq:Lvtr}\\
 (\BI{X},\BI{b},\BI{c})&\to&
	(\BI{X},e^{-\alpha_g}\BI{b},e^{\alpha_g}\BI{c}).\qquad\qquad
	(\,e^{\alpha_g}=(\bar{h}^2/h)^{1/4}\,)\label{eq:GA}
\end{eqnarray}
However, the rescalings in (\ref{eq:Lvtr}) and (\ref{eq:GA}) shall
generate the jacobians from the path integral measure \cite{F}.
We can see that (\ref{eq:Lvtr}) is a conformal
transformation and (\ref{eq:GA}) is a ghost number
transformation for the worldsheet reparametrization ghosts.
Since the superstring induced from the wrapped supermembrane is in the
critical dimension, the jacobian coming from (\ref{eq:Lvtr}) should be
trivial, or the conformal anomaly should be canceled.\footnote{Of
course, we should recover the fermionic coordinates and the bosonic
ghosts for the supertransformation in order to cancel out the
conformal anomaly (see section \ref{S:SD}).}
On the other hand, the jacobian from (\ref{eq:GA}) corresponds to the
reparametrization ghost number anomaly and it is well known that the
anomaly equation gives the local version of the Riemann-Roch
theorem,
\begin{equation}
 \p_{\bar{z}} j^{gh}_z = -\frac{3}{4\pi}\p_z\p_{\bar{z}}\ln\rho
	= \frac{3}{8\pi}\sqrt{\gstg^E}R^{(2)},
\end{equation}
where $j_z^{gh}$ is the worldsheet reparametrization ghost number
current and $R^{(2)}$ is the worldsheet curvature.
Thus, we have
\begin{eqnarray}
 \calD(e^{-\alpha_g}\BI{b})\calD(e^{\alpha_g}\BI{c})
  &=&\calD\BI{b}\calD\BI{c}\, J(\alpha_g)
	=\calD\BI{b}\calD\BI{c}\,e^{\ln J}\nn
 &=& \calD\BI{b}\calD\BI{c}\,\exp\Bigl[-\frac{3}{8\pi}
	\int\!d^2z\,\Phi(\alpha_g) \sqrt{\gstg^E}\,R^{(2)} \Bigr],
\end{eqnarray}
where
\begin{equation}
 \Phi(\alpha_g)=\ln\alpha_g
	=\frac{1}{4}\ln\left(\frac{\bar{h}^2}{h}\right)
	\simeq\frac{1}{4}\ln\hG_{1010}=\frac{1}{3}\,\phi\,,
\end{equation}
and $\phi$ is the background dilaton (cf. (\ref{eq:11BG})).
Then, we have (~$\BI{}$~  is omitted)
\begin{equation}
  \calD(e^{-\alpha_g}b_{zz})\calD(e^{-\alpha_g}b_{\bar{z}\bar{z}})
	\calD(e^{\alpha_g}c^z)\calD(e^{\alpha_g}c^{\bar{z}})
  =\calD b_{zz}\calD b_{\bar{z}\bar{z}}\calD c^z\calD c^{\bar{z}}\,
	\exp\Bigl[-\frac{1}{4\pi}\int\!d^2z \sqrt{\gstg^E}\,
	\phi R^{(2)}\Bigr].\label{eq:phi-R}
\end{equation}
We have seen that this dilaton-curvature term has appeared due to $h$,
which indicates that the term originated with the membrane.

\section{Discussion}\label{S:SD}
In this paper, we have studied the dilaton coupling in the string
theory, which is induced from a wrapped membrane.
We have seen that the dilaton coupling to the string worldsheet
curvature comes out from the path integral measure due to the ghost
number anomaly, which originates from the fact that the string is
reduced from the membrane.

There remains some points to be discussed here.
The induced string action (\ref{eq:Mac02a}) or (\ref{eq:Mac02aE})
has the conformal mode $\rho$ dependence.
This action is obtained from the Polyakov-type action (\ref{eq:Mac})
through the double dimensional reduction.
On the other hand, once we start with the Nambu-Goto action
(\ref{eq:Mac1}) and adopt essentially the same gauge conditions as
(\ref{eq:GF}),
\begin{equation}
  X^{10}=R_1\sigma^2,\quad
  \gstg^{X}_{01}=\gstg^{X}_{00}+\gstg^{X}_{11}=0,\label{eq:GFi}
\end{equation}
where $\gstg^{X}_{ij}$ is the induced metric (cf. (\ref{eq:IM}),
(\ref{eq:WV1A}))
\begin{equation}
 \gstg^{X}_{ij}= \elpl^{-1}\hG_{22}^{1/2}\hG_{ij}
	-\elpl^{-2}\hG_{22}^{-1}\hG_{i2}\hG_{j2}\,,\label{eq:IMij}
\end{equation}
the total action in this case becomes
\begin{eqnarray}
  S^{\mathrm{(NG)}}_T&=&\frac{T\elpl^3}{2}\int\!d^3\sigma\,\Biggl[
    -(\gstg_{00}^{X}- \gstg_{11}^{X})
    -\elpl^{-2}\epn^{ij}\p_{i}X^{\hmu}
	\p_{j}X^{\hnu}B_{\hmu\hnu}
	-\elpl^{-3}\epn^{ij}\p_{i}X^{\hmu}\p_{j}X^{\hnu}
	\p_{2}X^{\hrho}A_{\hmu\hnu\hrho}\nn
  &&{}\qquad-4i\hh^{1/2}\Bigl\{\bg_{++}(\p_--\hV_-\p_2)\cg^+
	+\bg_{--}(\p_+-\hV_+\p_2)\cg^-\Bigr\}\Biggr]\,.
\end{eqnarray}
We shall see that $S^{\mathrm{(NG)}}_T$ corresponds to $S_T$
(\ref{eq:MacT}) with $h=\hh$ and $V_i=\hV_i$.
Thus, the dilaton-curvature coupling term (\ref{eq:phi-R}) will be
followed by a similar analysis.

In this paper we have omitted the fermionic coordinates and hence
the worldvolume local supersymmetry.
However, we have used the fact that the jacobian coming from
(\ref{eq:Lvtr}) is trivial, which is to be explicitly shown with the
fermionic coordinates.
Since the wrapped supermembrane is reduced to the Green-Schwarz type
of superstring, but not the Neveu-Schwarz-Ramond type, such a  gauge
fixing as in \cite{B} will be desired. This should be studied further.

We have considered the $S^1$-compactification of the supermembrane.
In the meantime, it was shown explicitly that the wrapped
supermembrane on a 2-torus induces the ($p,q$)-string
\cite{OUY3,OUY}. Then, the dilaton-curvature coupling term in the type
IIB superstring would be directly presented with the wrapped
supermembrane.

We adopted the double dimensional reduction for the wrapped
supermembrane and did not investigate the Kaluza-Klein modes.
Meanwhile, quantum mechanical justifications of the double
dimensional reduction were studied in \cite{SY,UY}.
In order to study the coupling between $\rho$ and the
Kaluza-Klein modes, such quantum mechanical treatment should be
investigated further.

\vspace{\baselineskip}
\noindent{\bf Acknowledgments:}
The author would like to thank Hiroyuki Okagawa and Satoshi Yamada for
their collaboration at the early stage of this work.
This work is supported in part by MEXT Grant-in-Aid for Scientific
Research \#20540249.

\appendix
\section{11d vs. 10d background fields}\label{S:R}
The 11-dimensional metric can be written as
\begin{eqnarray}
 \hG_{MN}&\equiv& e^{-\frac{2}{3}\phi}
    \left(\begin{array}{@{\,}cc@{\,}}
	\ga_{\hmu\hnu}+e^{2\phi}A_{\hmu}A_{\hnu}&
		 e^{2\phi}  A_{\hmu} \\[10pt]
	e^{2\phi}  A_{\hnu} & e^{2\phi} \end{array}\right)\nn
 &=& \left(\begin{array}{@{\,}cc@{\,}}
	\frac{1}{\sqrt{\hG_{1010}}}\,\ga_{\hmu\hnu}
	+\frac{1}{\hG_{1010}}\hG_{\hmu10}\hG_{\hnu10}&
		 \hG_{\hmu10} \\[10pt]
	\hG_{\hnu10} & \hG_{1010} \end{array}\right),\label{eq:11BG}
\end{eqnarray}
and the third-rank antisymmetric tensor $\hA_{MNP}$ is decomposed as
\begin{equation}
 \{\hA_{MNP}\} = \{\hA_{\hmu\hnu\hrho},\hA_{\hmu\hnu10}\}
  =\{A_{\hmu\hnu\hrho},B_{\hmu\hnu}\}\,.\label{eq:dhA}
\end{equation}

\section{FP determinant (\ref{eq:MacFP})\label{S:FP02a}}
Let us calculate the FP determinant.
The gauge condition (\ref{eq:GF})
\begin{equation}
 X^{10}=R_1\,\sigma^2,\quad \gstg_{01}=\gstg_{00}+\gstg_{11}=0,
\end{equation}
leads to (cf. (\ref{eq:WV1A}))
\begin{eqnarray}
 \Delta_{\mathrm{FP}}(\hatg,X)^{-1}
 &=&\int\!\calD\hze\,\delta((X^{10})^{\hze}-R_1\,\sigma^2)\,
    \delta((\hatg_{\hi\hj}-\hat{\fdg}_{\hi\hj})^{\hze})\nn
 &=&\int\!\calD\hze\,\delta(R_1\hze^2)\,
	\delta(h^{-1/2}\delta_{\hze}(\Delta\gstg_{ij}))\nn
 &=&\int\!\calD\hze\,\delta(R_1\hze^2)\!
  \int\!\calD\hld\,\exp\Bigl[2\pi i\!\int\sqrt{-\hatg}\,
	h^{-1/2}\hld^{ij}\delta_{\hze}(\Delta\gstg_{ij})\Bigr]\nn
 &=&\int\!\calD\hze\,\delta(R_1\hze^2) \!\int\!\calD\hld\,\exp\Bigl[
    2\pi i\!\int \sqrt{-\fdg}\,h^{1/2}\,\check{\lambda}^{ij}
	\delta_{\hze}(\Delta\gstg_{ij})\Bigr],\label{eq:FP-01}
\end{eqnarray}
where $\hat{\fdg}_{\hi\hj}$ is given by substituting the fiducial
metric $\fdg_{ij}$ (\ref{eq:FDm}) for $\gstg_{ij}$ in
$\hatg_{\hi\hj}$, $\Delta\gstg_{ij}$ is the difference between
$\gstg_{ij}$ and $\fdg_{ij}$,
\begin{equation}
 \Delta\gstg_{ij}=\gstg_{ij}-\fdg_{ij},
\end{equation}
$\check{\lambda}^{ij}$ is traceless,
$\check{\lambda}^{ij}\fdg_{ij}=0$, and
$\delta_{\hze}(\Delta\gstg_{ij})=
\delta_{\hze}(\gstg_{ij}-\fdg_{ij})|_{\gstg_{ij}=\fdg_{ij}}$.
Thus,  eq.(\ref{eq:MacFP}) is followed by
\begin{eqnarray}
 \check{\lambda}^{ij}\delta_{\hze}(\Delta\gstg_{ij})
 &=&2\check{\lambda}^{01}\fdg_{00}(V_0\p_2\hze^1
	-V_1\p_2\hze^0+\p_1\hze^0-\p_0\hze^1)\nn
  &&{}+2\check{\lambda}^{00}\fdg_{00}(
	V_1\p_2\hze^1-V_0\p_2\hze^0+\p_0\hze^0-\p_1\hze^1)\nn
 &=&2\fdg^{11}\check{\lambda}_{01}
	(V_0\p_2\hze^1-V_1\p_2\hze^0+\p_1\hze^0-\p_0\hze^1)\nn
  &&{}+2\fdg^{00}\check{\lambda}_{00}(
	V_1\p_2\hze^1-V_0\p_2\hze^0+\p_0\hze^0-\p_1\hze^1)\nn
 &=&\frac{4}{\sqrt{-\det\fdg}}\Bigl\{
	\check{\lambda}_{++}(V_-\p_2-\p_-)\hze^+
	+\check{\lambda}_{--}(V_+\p_2-\p_+)\hze^-\Bigr\},
\end{eqnarray}
where
\begin{equation}
 \hze^{\pm}=\hze^0\pm\hze^1,\quad\p_{\pm}=\frac{1}{2}\,(\p_0\pm\p_1),
 \quad V_{\pm}=\frac{1}{2}\,(V_0\pm V_1),
 \quad\check{\lambda}_{\pm\pm}
    =\frac{1}{2}(\check{\lambda}_{00}\pm\check{\lambda}_{01}).
\end{equation}


\end{document}